\documentclass{article}
\usepackage{eurosym}
\usepackage{amsmath}
\usepackage{amsfonts}
\usepackage{graphicx,epsfig}
\usepackage{epstopdf}
\usepackage{amssymb}

\begin{document}

\title{\textbf{\LARGE Bouncing cosmology in $f(R,T)$
gravity}}
\author{{\large J. K. Singh}$^{1}$, {\large Kazuharu Bamba}$^{2}$, {\large Ritika Nagpal}$^{3}$, {\large S.
K. J. Pacif}$^{4}$ \\ \\
$^{1,3}$\textit{Department of Mathematics,}\\
\textit{\ \ Netaji Subhas Institute of Technology,}\\
\textit{\ \ Faculty of Technology, University of Delhi,}\\
\textit{\ \ New Delhi 110 078, India}\\ \\
$^{2}$\textit{Division of Human Support System,}\\
\textit{\ \ Faculty of Symbiotic Systems Science,}\\
\textit{\ \ Fukushima University, Fukushima 960-1296, Japan}\\ \\
$^{4}$\textit{Centre for Theoretical Physics,}\\
\textit{\ Jamia Millia Islamia,\ }\\
\textit{New Delhi 110 025, India}\\ \\
\textit{\ \ jainendrrakumar@rediffmail.com}$^{1}$, \textit{bamba@sss.fukushima-u.ac.jp}$^{2}$\\ \\ \textit{
ritikanagpal.math@gmail.com}$^{3}$, \textit{shibesh.math@gmail.com}$^{4}$\\
}
\maketitle

{\footnotesize \vskip0.4in \noindent \textbf{Abstract} A cosmological model with a specific form of the Hubble parameter is constructed in a flat homogeneous, and isotropic background in the framework of $f(R, T)$ gravity, where $R$ is the scalar curvature and $T$ is the trace of the stress-energy-momentum tensor. The proposed functional form of the Hubble parameter is taken in such a way that it fulfills the successful bouncing criteria to find the solution of the gravitational field equations provided the Universe is free from initial singularity. The various constraints on the parameters are involved in the functional form of the Hubble parameter which is analyzed in detail. In addition, we explore physical and geometrical consequences of the model based on the imposed constraints. Furthermore, we demonstrate the bouncing scenario which are realized in our model with some particular values of the model parameters. As a result, we find that all of the necessary conditions are satisfied for a successful bouncing model.
}

{\footnotesize \vskip0.2in \noindent \textbf{Keywords}\,\, f(R,T) gravity $\cdot $ Hubble Parameter $\cdot $ Bouncing model $\cdot $ Scalar field \newline
 }

\section{\protect\Large Introduction}

\qquad In the present scenario, we are familiar with the accelerating expansion of the Universe in the early phase of evolution (inflation) as well as in the late Universe. Various cosmological observations give evidence of late time cosmic acceleration. Some of them are observations of high red-shift supernovae \cite{rie}, cosmic microwave background radiations (CMBR) \cite{spe1,spe2}, supernovae of type Ia \cite{per,benn}, Planck data \cite{ade} and baryon acoustic oscillations \cite{perc}. The fundamental pillar of the modern cosmology is the general theory of relativity (GTR). As of now GTR explains the large-scale structure of the Universe very well theoretically. In order to explain the late-time cosmic acceleration, different approaches have been developed in the past few decades \textit{e.g.} modifying the energy momentum tensor or modifying the geometry in the Einstein's field equations (EFEs). The inclusion of the matter with highly negative pressure termed as dark energy (DE) in the right side of EFE is much successful in explaining the puzzle of the late-time Universe as observations reveal. But the nature of this mysterious DE is unknown to us as there is no any direct evidence of DE. Generally, it is  believed that DE is a homogeneous fluid that permeates all over the space contributing almost 2/3 of the total volume of the Universe. However, this is matter of a great debate on the candidature of DE among the theorists. \newline

\qquad Many theoretical cosmological models of the Universe have been suggested to examine the behavior of the DE. Out of the numerous candidates of DE, the most efficient and prospective version is the cosmological constant introduced by Einstein \cite{star, pee}. However, it is afflicted by most familiar cosmological constant problem \cite{wein} which can be soothed by assuming a dynamical decaying $(\Lambda )$. In another way, early inflation plays a major role in understanding the anisotropies in the CMBR and formation of large scale structures. Literature includes the list of DE candidates, such as quintessence, f-essence, k-essence, spintessence, tachyons, phantom, Chaplygin gas \textit{etc.} (for a brief review see \cite{SAMI}). Despite of these excellent DE cosmological models, we lack behind on some issues and the hunt for a concrete model is still open. This motivates the theorists to consider alternative theories of gravity. The modification of GTR in which the origin of DE is associated by the rearrangement in gravity as it is described by Riemannian geometry and is without torsion. Although in many literatures, a number of gravitational theories have been studied in which the torsion effects appear in the extension of GTR. Some of the alternating theories are $f(R)$ gravity \cite{buc, noj, sot, cap, cli}, $f(T)$ gravity \cite{fem}, where $T$ is the torsion scalar in teleparallel gravity, $ f(G) $ gravity \cite{tre}, where $G$ is the Gauss-Bonnet invariant, \textit{etc.} (for recent reviews on the dark energy problem and modified gravity theories, see \cite{bamba1}). However, many alternative theories have been studied in the past few decades such as \text{Brans-Dicke} theory, Einstein-Cartan theory, Loop quantum theory, \text{ Kaluza-Klein} theory and many more to overcome the cosmological issues.\\

\qquad Another prospective and efficient theory among the alternative theories is $f(R,T)$ gravity, where $ R $ and $ T $ are the Ricci scalar curvature and trace of stress-energy-momentum tensor (SEMT) respectively \cite{tfs}. According to this theory there is an arbitrary coupling constant between matter and geometry which is responsible for a source term, performing the matter-stress-energy tensor \textit{with respect to} the metric. The different choice of the matter Lagrangian $L_{m}$ would cause an explicit set of field equations. So many new investigations have come on the surface to describe the present cosmic acceleration in $f(R,T)$ gravity  \cite{sin1, sin2, sah, nat, fara, alve, lad, sin3, prad, zub, buf}. Several homogeneous isotropic and anisotropic cosmological models have been constructed in $f(R,T)$ theory of gravity for the past few years \cite{sin4, sha, sin5, nou, bha, sin6, alv, mor, bani}. Houndjo \textit{et al.} \cite{Houndjo:2011fb} has reconstructed a cosmological model in $ f(R, T) $ gravity which is able to discuss the expansion history of the model in GTR by dark matter as well as by holographic dark energy. Barrientos \textit{et al.} \cite{Barrientos:2018cnx} has studied the $ f(R,T) $ theories of gravity and its application with affine connection.\\

\qquad According to the standard cosmological model, the Universe came into existence from a singularity (big bang) in the space-times which has some shortcomings \textit{e.g.} horizon problem, flatness problem, transplanckian problem, entropy problem, original structure problem, and singularity problem. To resolve these issues, a sudden growth after the big bang was necessary to constitute a uniform, flat and smooth universe. Inflationary theory was developed by Alan Guth to solve a bunch of these standard cosmological problems which proved to be a quite successful to describe the various observational properties of the Universe. However, one of the fundamental questions in modern cosmology still remains unsolved \textit{i.e.} the initial singularity problem. One of the attractive possible alternatives to the inflationary model was developed as a bouncing model of the Universe which solves the initial singularity problem during mid 1980's. The specification of this bouncing scenario is that the Universe may have emerged from a prior contracting phase and capable of being expanding without singularity or it experiences a bouncing process \cite{shab, kuir}. 
Some excellent research works on a nonsingular bouncing cosmology have been carried out by Cai \textit{et al.} \cite{cai1,cai2,cai3,cai4,cai5} and Brandenberger \textit{et al.} \cite{bran} in recent years in which they studied the various phenomenological aspects of the bouncing scenario\textit{ e.g.} the cosmology of a Lee-Wick type scalar field theory, single scalar field matter containing a potential and kinetic term, a contracting universe that consists of radiation, cold dark matter (CDM) and a positive cosmological constant, bounce model with dark matter (DM) and DE, observational bouncing cosmologies such as the Planck and BICEP2 data, and the role of bouncing cosmologies as alternative theories to the cosmological inflation which are consistent with present day observational data.\\

\qquad Bamba \textit{et al.} \cite{bamba2, bamba3, bamba4} have discussed bouncing cosmological models in $ f(R)$ gravity by reconstructing a method, in Gauss-Bonnet gravity where the Gauss-Bonnet invariant couples to a dynamical scalar field, in $ f(G) $ gravity with the Gauss-Bonnet invariant $G$ by reconstructing a method to search the bouncing scenario in the early Universe as well as  examine the stability conditions for its solutions, and a bouncing inflationary model with a graceful exit into the Friedmann-Lemaitre-Robertson-Walker (FLRW) model in $ f(T) $ gravity, $ T $ being the torsion scalar respectively. Bamba \textit{et al.} \cite{bamba5} have also explored a bouncing inflationary model with a graceful exit into the FLRW model in $ f(T) $ gravity. de la Cruz-Dombriz \textit{et al.} \cite{delaCruz-Dombriz:2018nvt} has discussed the bouncing cosmological model in the extended theory of teleparallel gravity. Cai \textit{et al.} \cite{cai6} have examined the matter bounce cosmological models in $ f(T) $ gravity.\\ 

\qquad In this work, we propose a new form of the parametrization of HP which varies with cosmic time $t$ and study the evolution of the Universe in $ f(R,T) $ gravity in the framework of a flat FLRW metric. Different cases arise while imposing the restrictions on model parameters, which involved the functional form of the Hubble parameter $ H $. This leads to some accelerating expansion without singularity in a bouncing scenario. The physical consequences of the model have also been discussed.\\

\qquad The outline of the work is as follows: In Sect. II, we present a brief review on $ f(R,T) $ gravity and discuss the metric and its field equations. In Sect. III, we study the necessary conditions in order to accomplish a successful bouncing model of the Universe in standard cosmology and discuss the Hubble parameter (HP), Deceleration parameter (DP), equation of state (EoS) parameter for a bouncing model, and analyze the HP and DP mathematically on various constraints. In Sect. IV, we take a new parametrization of HP for some specific values of the model parameters $ c $ and $ \lambda _{2} $ to understand our proposed model through various plots. In Sect. V, we examine the scalar field and self interacting potential by adopting Barrow's scheme \cite{barr}. Finally, we discuss and summarize the results in section VI.\\

\section{OVERVIEW OF ${\protect\Large f(R,T)}${\protect\Large \ GRAVITY AND FIELD EQUATIONS }}
\qquad The $f(R,T)$ theory is the modification of GTR in which a matter Lagrangian $L_m$ can be described by the combination of $R$ and $T$,  where $R$ and $T$ are the scalar \cite{tfs}. The total
gravitational action of $f(R,T)$ gravity becomes

\begin{equation}\label{1}
S=\frac{1}{16\pi G}\int f(R,T)\sqrt{-g}d^{4}x+\int L_{m}\sqrt{-g}d^{4}x,
\end{equation}
where $g$ and $G$ indicate the metric determinant and the gravitational constant respectively. Several forms of $f(R,T)$ function are given in literature; here we are assuming the coupling between $R$ and $T$ in the form $f(R,T)=R+2f(T)$. Furthermore, we assume $ f(T)=\lambda_{1}T $, where $ \lambda _{1} $ is an arbitrary constant. The above functional form of $ f(R,T) $ is designed in such a manner that GTR can be obtained for $\lambda_1=0$. By varying the action (\ref{1}) \textit{with respect to} the metric tensor components yield

\begin{equation} \label{2}
G_{ij}+\left( g_{ij}\square -\nabla _{i}\nabla _{j}\right) =[8\pi
+2f^{\prime }(T)]T_{ij}+2[f^{\prime }(T)p+\frac{1}{2}f(T)]g_{ij},
\end{equation}%
\newline
where prime denotes differentiation \textit{with respect to} the argument. We consider, here, a perfect fluid as the matter source filled in the Universe and therefore, SEMT of $L_m$ can be taken as
\begin{equation}\label{3}
T_{ij}=(\rho +p)u_{i}u_{j}-pg_{ij},
\end{equation}
\newline
where $\rho$ and $p$ are the cosmic energy density and isotropic pressure of the fluid respectively, and $u^{i}=(0,0,0,1)$ represents the four velocity vector components in the comoving coordinate system which satisfies the conditions $ u^{i}u_{i}=1 $ and $u^{i}\nabla _{j}u_{i}=0$. We choose the perfect fluid
matter as $L_{m}=-p$ in the action (\ref{1}).\\

We consider a flat FLRW background geometry with a metric
\begin{equation} \label{4}
ds^{2}=dt^{2}-a^{2}(t)\sum_{i=1}^{3}dx_{i}^{2},
\end{equation}%
where $a(t)$ is the scale factor; the gravitational field equations in $f(R,T)$ gravity (\ref{2}) are  
\begin{equation}\label{5}
3H^{2}=(8\pi +3\lambda _{1})\rho -\lambda _{1}p,
\end{equation}
\begin{equation} \label{6}
2\dot{H}+3H^{2}=-(8\pi +3\lambda _{1})p+\lambda _{1}\rho ,
\end{equation}
\newline
where an overhead dot indicates the differentiation of the quantity \textit{with respect to} cosmic time $ t $. In the present scenario, we need one more constraint equation to solve the system of field equations completely. This constraint equation is the state equation of a fluid in general. But, here we are interested in a bouncing scenario which is not a new concept and has a long history for which we put a constraint on the  Hubble parameter \cite{dia, anna}. However, one can also consider a parametrization of physical or geometrical parameters to get a constraint equation which is consistent with the system. This is generally called the \textit{model independent way} to study the DE models without violating the background theory. For a brief review for various parametrizations of cosmological parameters, see \cite{cai7,cai8,SKJP}.

\section{\protect\Large BOUNCING SOLUTION AND PARAMETRIZATION OF $ H $}

\qquad In recent times, the bouncing scenario has gained popularity where the big bang is replaced by big bounce. The big bounce cosmological model can be interpreted as an oscillatory universe or the cyclic universe where one cosmological event was the outcome of the collapse of another or previous universe. A bouncing universe which contracted to a finite volume initially, then expanded subsequently provides a possible solution to the singularity problem of the standard big bang theory within GTR. For a successful bounce, it can be observed that the violation of a null energy condition (NEC) is required for a period of time in the vicinity of the bouncing point within a FLRW background geometry.  Moreover, the EoS parameter $\omega$ of the matter content present in the Universe must undergo a phase transition from $ \omega< -1 $  to $ \omega> -1 $, to enter into the hot big bang age after the bounce \cite{cai7,cai8}. The observational data \cite{zho} support the quintom model \cite{fen}, which has been proposed to explore the behavior of the DE with an EoS parameter $ w > -1$ and  $w < -1$  in the past and at present respectively. The quintom model is a nonstatic model of DE which behaves distinctively from the other DE models \textit{e.g.} cosmological constant, quintessence, phantom, k-essence \textit{etc.} in the determination of the cosmic evolution.\\ 

The detailed description on the necessary conditions in order to accomplish a successful bouncing model of the Universe in standard cosmology are as follows \cite{cai7}:

\begin{itemize}

\item In the contracting universe, a scale factor $a(t)$ is decreasing \textit{i.e.} $ \dot{a(t)} < 0 $, and in the expanding universe, the scale factor $ a(t) $ is increasing, \textit{i.e.} $ \dot{a(t)} > 0 $. The cosmic scale factor reaches to a nonzero minimum value at the transfer point. This kind of bouncing scenario can avoid the singularity naturally which is inevitable in the standard model. In other words, during the bouncing point, $ \dot{a}(t)=0 $ and $ \ddot{a}(t)>0 $ for some period of time in the neighborhood of a bounce point.

\item  Equivalently the HP passes through zero from $ H < 0 $ when the Universe contracts to $ H > 0 $ when the Universe expands and $ H = 0 $ when the bouncing point occurs. A successful bouncing model in standard cosmology requires $ \dot{H}= −4\pi G \rho(1 + w) > 0 $ in the neighborhood of a bouncing point which is equivalent to the violation of the null energy condition (NEC) in Einstein gravity.  From this equation, one can observe that $ w < -1 $ around the bouncing point.                      

\item The EoS parameter $ \omega $ crosses the phantom divide line (quintom line) $ \omega=-1 $ which is the remarkable feature of the quintom model.
\end{itemize}

Motivated by the above bouncing scenario and the model independent way to study cosmological models, here in this paper, we would like to emphasize on the cosmographic parameter $ H $ that describes the expansion of the Universe and helps us to achieve some impressive bouncing solutions to the EFEs. We parametrize the functional form of the Hubble parameter as a product function given by
\begin{equation}\label{7}
H(t)=\alpha \,t\,h(t),  
\end{equation}%
where $ \alpha $ is any arbitrary constant and $h(t)$ is any analytic function. Looking at the proposed form of the Hubble parameter, we can observe that the algebraic function $t$ present in the functional forces to vanish $ H $ at $ t=0 $ implying that the scale factor function must take a constant value at $ t=0 $ [the second function $ h(t) $ being nonvanishing at $ t=0 $]. We have a choice on $ h(t) $ which could be any rational or transcendental, or periodic function to get another bounce in the future. Here, in this study, we consider a specific functional form of $ h(t) $ as
\begin{equation}
h(t)=ln\Big(\frac{c-\lambda _{2}\tan ^{-1}t}{t}\Big),  \label{8}
\end{equation}%
so the parametrization of HP takes the form
\begin{equation}
H(t)=\alpha \,t\,ln\Big(\frac{c-\lambda _{2}\tan ^{-1}t}{t}\Big),  \label{9}
\end{equation}%
$\allowbreak $ where $\lambda _{2}$ and $ c $($>0$) are arbitrary constants and have time dimensions (we call them model parameters), which describe the dynamics of the Universe. In the following sections, we discuss the cosmological parameters for our model.

\subsection{Hubble parameter}

\noindent \qquad The aforesaid form of the HP in Eq. (\ref{9}) is bouncing in nature and have some specific features. The sign of Hubble parameter decides the expansion, contraction and bounce of the Universe. In our case

\begin{itemize}
\item The Universe is expanding if both $\alpha $ and $h(t)$ have same signature.

\item The Universe is contracting if $\alpha $ and $h(t)$ have opposite signature.

\item The Universe bounce \textit{i.e.} $H=0$ in the case either $\alpha $
or $ h(t)$ vanishes. Here, bounce occurs at $t=0$ (initial bounce) and $ t=\frac{3(1+\lambda _{2})2^{\frac{1}{3}}}{\left( -81c\lambda _{2}^{2}+\sqrt{6561c^{2}\lambda _{2}^{4}-2916\lambda _{2}^{3}(1+\lambda _{2})^{3}}\right) ^{
\frac{1}{3}}}+\frac{\left( -81c\lambda _{2}^{2}+\sqrt{6561c^{2}\lambda_{2}^{4}-2916\lambda _{2}^{3}(1+\lambda _{2})^{3}}\right) ^{\frac{1}{3}}}{3\lambda _{2}2^{\frac{1}{3}}} $(approx.) (future bounce).
\end{itemize}

We can discuss the future bounce \textit{w.r.t.} $\lambda _{2}$ and $ c $ through the following plots shown in Figs. 1a and 1b.
\begin{figure}[tbph]
\begin{center}
$%
\begin{array}{c@{\hspace{.1in}}cc}
\includegraphics[width=3in]{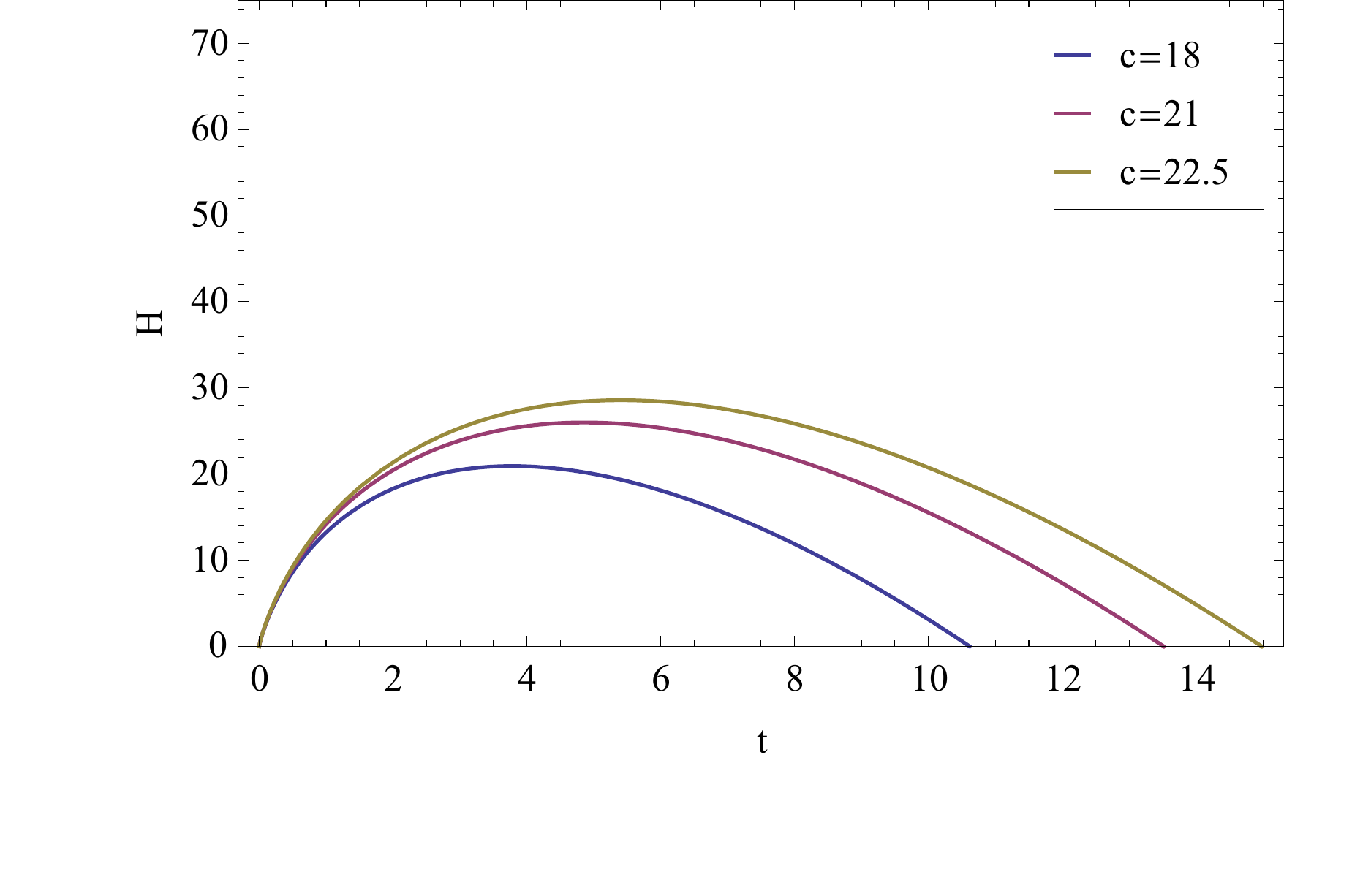} & \includegraphics[width=3in]{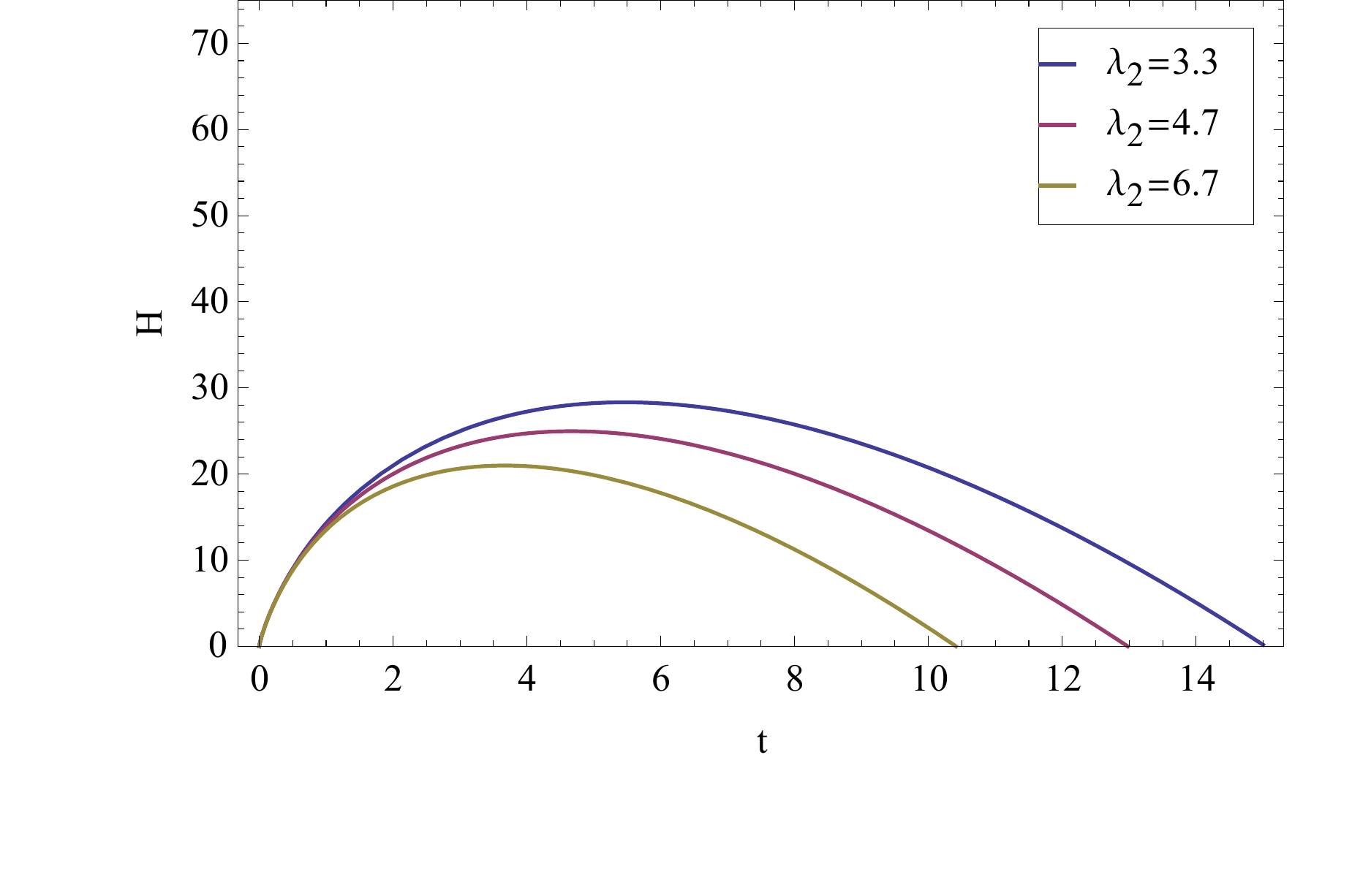}
&  \\
\mbox (a) & \mbox (b) &
\end{array}%
$%
\end{center}
\caption{ {\protect\scriptsize (a) The plot of Hubble parameter $H$ \textit{vs.} $t$ with fixed values of $\protect\alpha $, $\protect\lambda _{2}$, and different values of parameter $c$. (b) The plot of Hubble parameter $H$ \textit{vs.} $t$ with fixed values of $\protect\alpha $, $c$, and different values of parameter $\protect\lambda _{2}$.}}
\end{figure}
In these figures, we have fixed $\alpha =5$ ($\alpha $ can be treated as scaling constant) and plotted $H(t)$ for different values of model parameter $c$ by taking a fixed value of $\lambda _{2}$ in Fig. 1a. Similarly, we have fixed the value of $c$ in Fig. 1b and plotted $ H(t) $ for different values of $\lambda _{2}$. We observe that the future bounce depend on the value of model parameters $c$ and $\lambda _{2}$. The future bounce is delayed by increasing the value of $c$ (see Fig. 1a). Similarly the future bounce is delayed by decreasing the value of $\lambda _{2}$ (see Fig. 1b). This behavior of future bounce is analyzed with the constrain $ c>t>0 $
(see case 1 of Table 1).\newline

The different inequalities on the model parameters $ \alpha $ , $ \lambda _{2} $, $ c $ leading to some cases of expansion of the Universe are given in the following Table 1.

\begin{center}
{\scriptsize \
\begin{tabular}{|c|c|c|c|}
\multicolumn{4}{c}{\textbf{Table 1.} \textbf{Inequalities on model parameters in case of expansion}} \\ \hline
{\small cases} & {\small sign of }${\small \alpha }$ & {\small constraints
on }${\small \lambda }_{2}$ & {\small constraints on }${\small c}$ \\ \hline
${\small 1}$ & ${\small >0}$ & ${\small \lambda }_{2}{\small <}\frac{{\small %
c-t}}{\tan ^{-1}t}$ & $c>t>{\small 0}$ \\
${\small 2}$ & ${\small >0}$ & ${\small \lambda }_{2}{\small <}\frac{{\small %
c-t}}{\tan ^{-1}t}$ & ${\small 0<c<t}$ \\
${\small 3}$ & ${\small >0}$ & ${\small \lambda }_{2}{\small <0}$ & ${\small %
0<t<c}$ \\
${\small 4}$ & ${\small <0}$ & ${\small \lambda }_{2}{\small >}\frac{{\small %
c-t}}{\tan ^{-1}t}$ & ${\small c>t>0}$ \\
${\small 5}$ & ${\small <0}$ & $\frac{{\small c-t}}{\tan ^{-1}t}{\small %
<\lambda }_{2}{\small <0}$ & ${\small 0<c<t}$ \\ \hline
\end{tabular}%
}
\end{center}

\subsection{Deceleration parameter}

\qquad The deceleration parameter $q$ is given by
\begin{equation}\label{10}
q=-\frac{\ddot{a}a}{\dot{a}^{2}}=-1-\frac{\dot{H}}{H^{2}}.
\end{equation}
Using the functional form of $H(t)$ into Eq. (\ref{10}), the expression for the deceleration parameter $q$ is obtained as
\begin{eqnarray}\label{11}
q &=&-1+\frac{1}{\alpha t^{2}log\Big(\frac{c-\lambda_2 \tan ^{-1}t}{t}\Big)^{2}%
}\Big[1+\frac{\lambda_2 t}{(1+t^{2})(c-\lambda_2 \tan ^{-1}t)}  \notag \\
&&-log\Big(\frac{c-\lambda_2 \tan ^{-1}t}{t}\Big)\Big],
\end{eqnarray}
which depends on cosmic time $t$, and the signature of $q$ depends on the model parameters that describes the dynamics of the Universe. The Universe exhibits accelerated expansion according as the current observations. Hence, we emphasize on accelerating cases only. The constraints on model parameters for the accelerating cases is estimated in Table 2.
\begin{center}
{\scriptsize \
\begin{tabular}{|c|c|c|c|}
\multicolumn{4}{c}{\textbf{Table 2.} \textbf{Inequalities on model parameters in case of eternal acceleration}} \\ \hline
{\small cases} & {\small sign of }${\small \alpha }$ & {\small constraints
on }${\small \lambda }_{2}$ & {\small constraints on }${\small c}$ \\ \hline
{\small 1} & ${\small <0}$ & $\frac{{\small c-t}}{\tan ^{-1}t}<{\small %
\lambda }_{2}<\frac{{\small c}}{\tan ^{-1}t}$ & ${\small c>t>0}$ \\
{\small 2} & ${\small <0}$ & $0<{\small \lambda }_{2}<\frac{{\small c}}{\tan
^{-1}t}$ & ${\small 0<c<t}$ \\
{\small 3} & ${\small >0}$ & ${\small \lambda }_{2}<\frac{{\small c-et}}{%
\tan ^{-1}t}$ & ${\small 0<c<et}$ \\
{\small 4} & ${\small >0}$ & ${\small \lambda }_{2}<0$ & ${\small c>et}$ , where $e=2.7182$ (approx.)\\
\hline
\end{tabular}%
}
\end{center}

\begin{figure}[tbph]
\begin{center}
\includegraphics[width=3in]{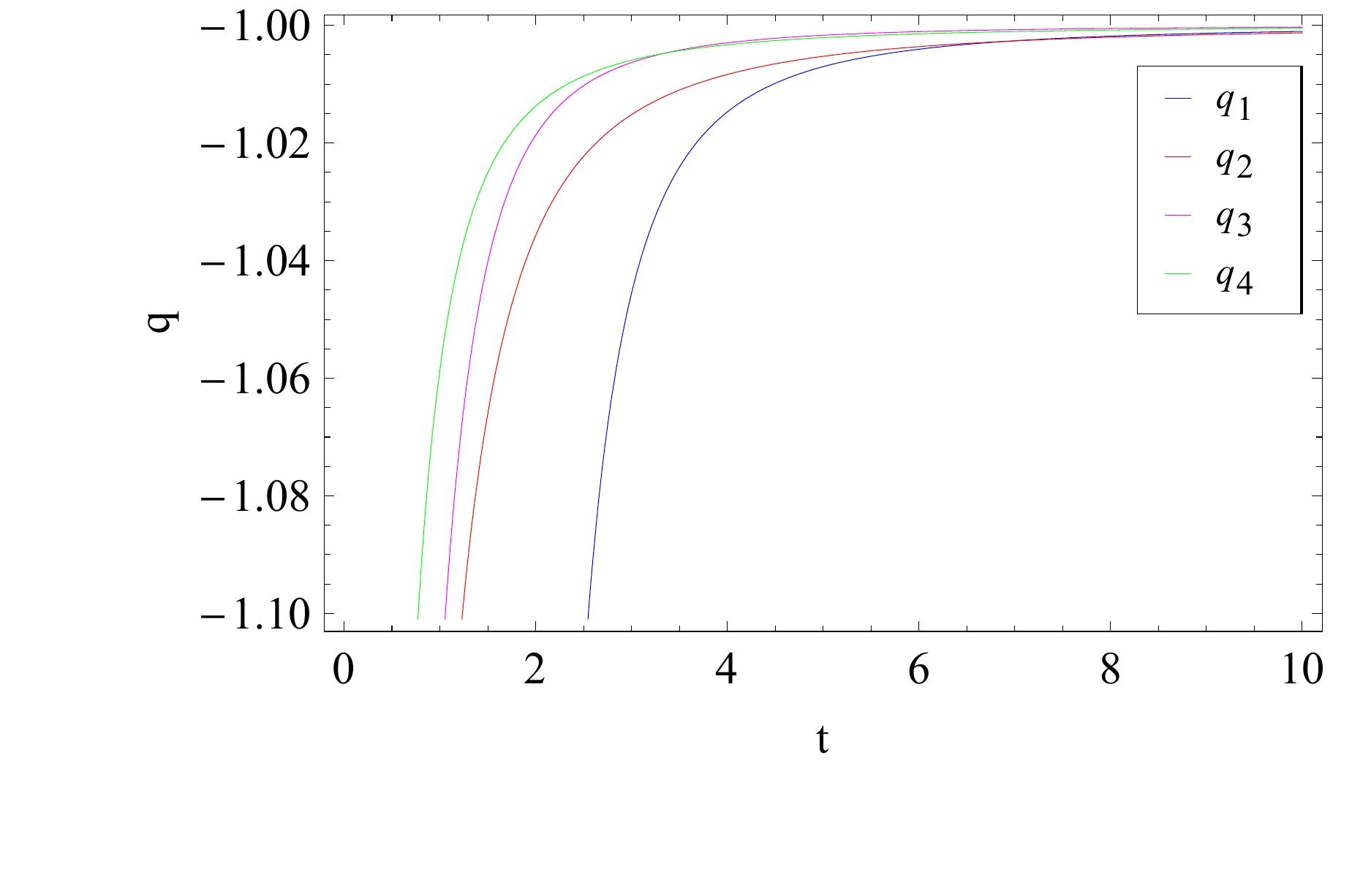}
\end{center}
\caption{ {\protect\scriptsize The plot of q \textit{vs.} $t $ of the
Universe, here $q_{i}$ indicates the case $i$, where $i=1,2,3,4 $ in Table 2.
}}
\end{figure}

From the plot of deceleration parameter (see Fig. 2), we can easily inspect that the Universe exhibits acceleration throughout the evolution under the restrictions on the model parameters given in Table 2.

\subsection{EoS parameter}

\qquad Here, we have two constants $\lambda _{1}$ and $\lambda _{2}$ involved in our model. Without the loss of generality, we assume that $\lambda _{1}=\lambda _{2}=\lambda $ to get a deterministic solution. With the help of assumption of Hubble parameter given in Eq. (\ref{9}), the EFEs (\ref{5}) and (\ref{6}) can now be solved explicitly. The expressions for the energy density $\rho $ and the pressure $p$ of the cosmic fluid are
\begin{eqnarray} \label{12}
\rho &=&\frac{\alpha }{4(4\pi +\lambda )(2\pi +\lambda )}\Big[\lambda +\frac{%
\lambda ^{2}t}{(1+t^{2})(c-\lambda tan^{-1}t)}-\lambda \log \big(\frac{%
c-\lambda tan^{-1}t}{t}\big)  \notag \\
&&+3\alpha (4\pi +\lambda )t^{2}\log \big(\frac{c-\lambda tan^{-1}t}{t}\big)%
^{2}\Big],
\end{eqnarray}

\begin{eqnarray}\label{13}
p &=&\frac{\alpha }{4(2\pi +\lambda )}\Big[\frac{(8\pi +3\lambda
)(c+ct^{2}+\lambda t-(1+t^{2})\lambda tan^{-1}t}{(1+t^{2})(4\pi +\lambda
)(c-\lambda tan^{-1}t)}  \notag \\
&&-\frac{(8\pi +3\lambda )log(\frac{c-\lambda tan^{-1}t}{t})}{4\pi +\lambda }%
-3\alpha t^{2}\,log\big(\frac{c-\lambda tan^{-1}t}{t}\big)^{2}\Big].
\end{eqnarray}
The EoS parameter is 

\begin{equation}\label{14}
\omega =\frac{\frac{\alpha }{4(4\pi +\lambda )(2\pi +\lambda )}\Big[\lambda +%
\frac{\lambda ^{2}t}{(1+t^{2})(c-\lambda tan^{-1}t)}-\lambda \log \big(\frac{%
c-\lambda tan^{-1}t}{t}\big)+3\alpha (4\pi +\lambda )t^{2}\log \big(\frac{%
c-\lambda tan^{-1}t}{t}\big)^{2}\Big]}{\frac{\alpha }{4(2\pi +\lambda )}\Big[%
\frac{(8\pi +3\lambda )(c+ct^{2}+\lambda t-(1+t^{2})\lambda tan^{-1}t}{%
(1+t^{2})(4\pi +\lambda )(c-\lambda tan^{-1}t)}-\frac{(8\pi +3\lambda )log(%
\frac{c-\lambda tan^{-1}t}{t})}{4\pi +\lambda }-3\alpha t^{2}\,log\big(\frac{%
c-\lambda tan^{-1}t}{t}\big)^{2}\Big]},
\end{equation}%
whose behavior can be analyzed on the various constrains given in Table 1 (see Fig. 3).

\begin{figure}[tbph]
\begin{center}
\includegraphics[width=3in]{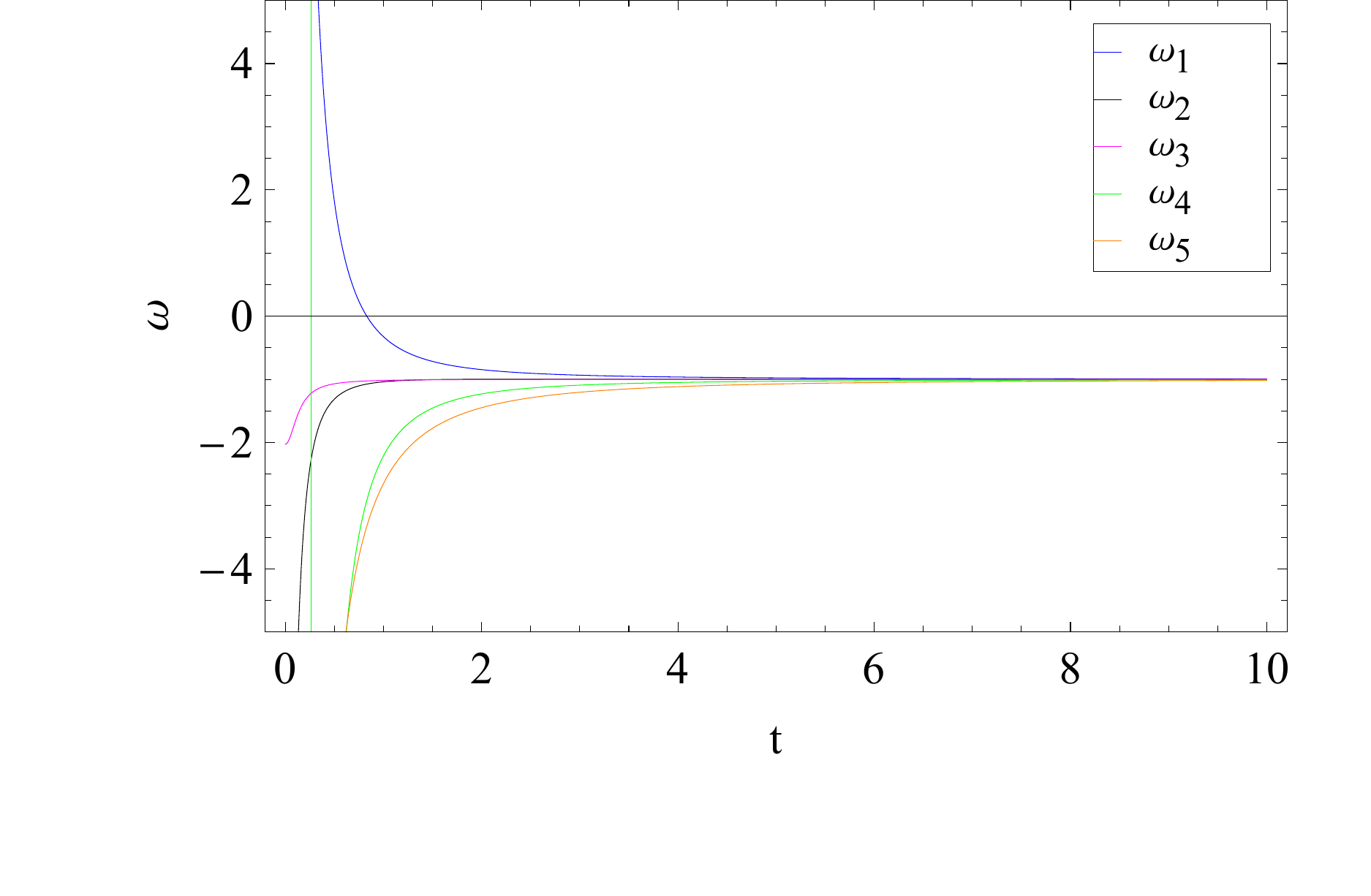}
\end{center}
\caption{{\protect\scriptsize The plot of $\protect\omega $ \textit{vs.} $t$%
. Here $\protect\omega _{i}$ indicates the case $i$, where $i=1,2,3,4,5$ as
in Table 1.}}
\end{figure}

In case 1 of Table 1, when the coupling constant of $f(R,T)$ gravity $\lambda $ is positive, the energy density $\rho $ and the pressure $p$ satisfies the EoS parameter $\omega $ with $\omega \geq 0$ at initial stage. In the late time $\omega $ transits its phase from perfect fluid to the quintessence region $-1<\omega <0$ and approaches to the phantom divide line (quintom line) but never enters in to the phantom region. In case 2, 3 and 5 of Table 1, energy density and pressure satisfies the EoS $\omega <-1$ due to the negative coupling constant of $f(R,T)$ gravity and in the late time approaches to the quintom line $ \omega =-1 $. In case 4 of Table 1, EoS parameter decreases promptly from perfect fluid region to phantom phase, and approaches to quintom line in late times.

\section{Exemplification}

\qquad We take some particular values of the model parameters $c$ and $\lambda _{2}$ to have a concrete understanding of our proposed model. We consider a more concise form of the function $h(t)$ given in Eq. (\ref{8}) by restricting the term $ tan^{-1}t $ up to third order only and by taking the model parameter $ c=1$ and $ \lambda _{2}=1 $. The new parametrization of HP in Eq. (\ref{9}) takes the specific form as
\begin{equation} \label{15}
H(t)=\alpha \,t\,ln\Big(\frac{1-t+t^{3}}{t}\Big),
\end{equation}
which is bouncing in nature having bounce at $ t\simeq0.618 $ (we have neglected the coefficient of $t^{3}$ for mathematical ease).

\begin{figure}[tph]
\begin{center}
{\scriptsize $%
\begin{array}{c@{\hspace{.1in}}cc}
\includegraphics[width=3in]{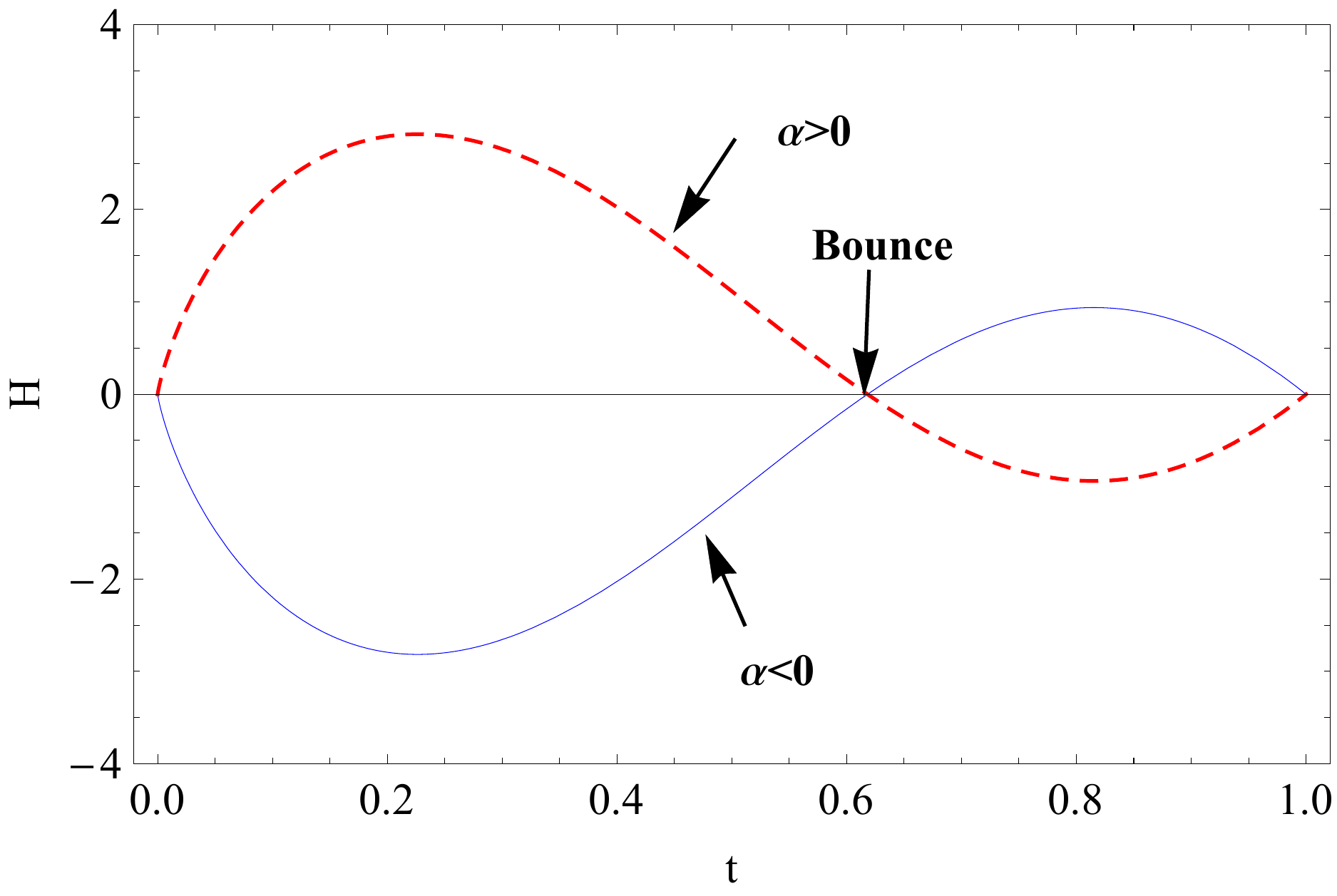} & \includegraphics[width=3in]{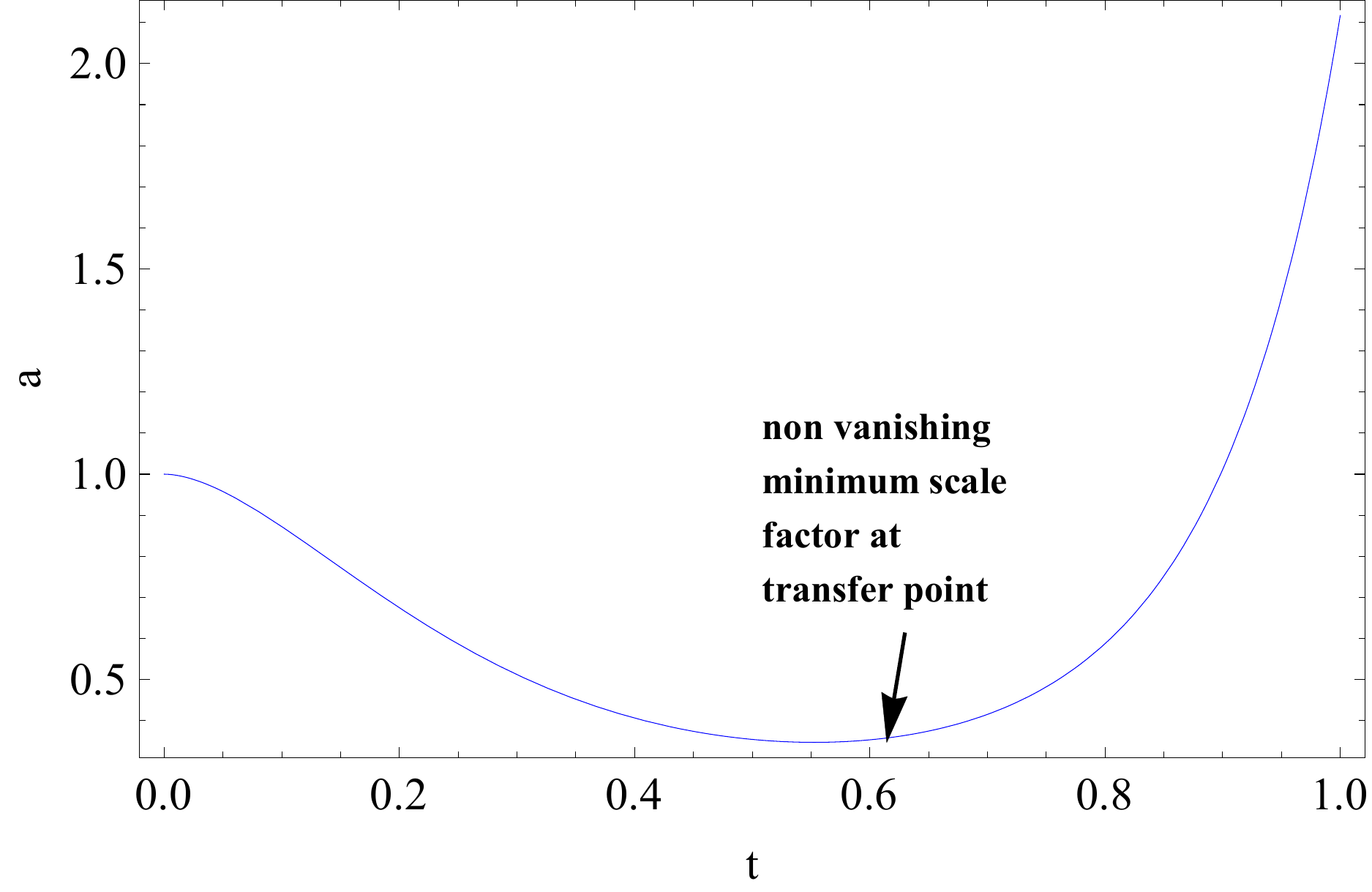}
&  \\
\mbox (a) & \mbox (b) &
\end{array}%
$ }
\end{center}
\caption{ \scriptsize (a) The plot of Hubble parameter $H$ \textit{vs.} $t$ for $\protect \alpha =-10$ (blue line) that describe contraction to expansion and $\protect\alpha =+10$ (red dashed line) that describe expansion to contraction. (b) The evolution of scale factor $a(t)$ \textit{vs.} $t$ for $\protect\alpha=-10$ which is for contracting to expanding Universe only.}
\end{figure}

\begin{figure}[tph]
\begin{center}
{\scriptsize $%
\begin{array}{c@{\hspace{.1in}}cc}
\includegraphics[width=3in]{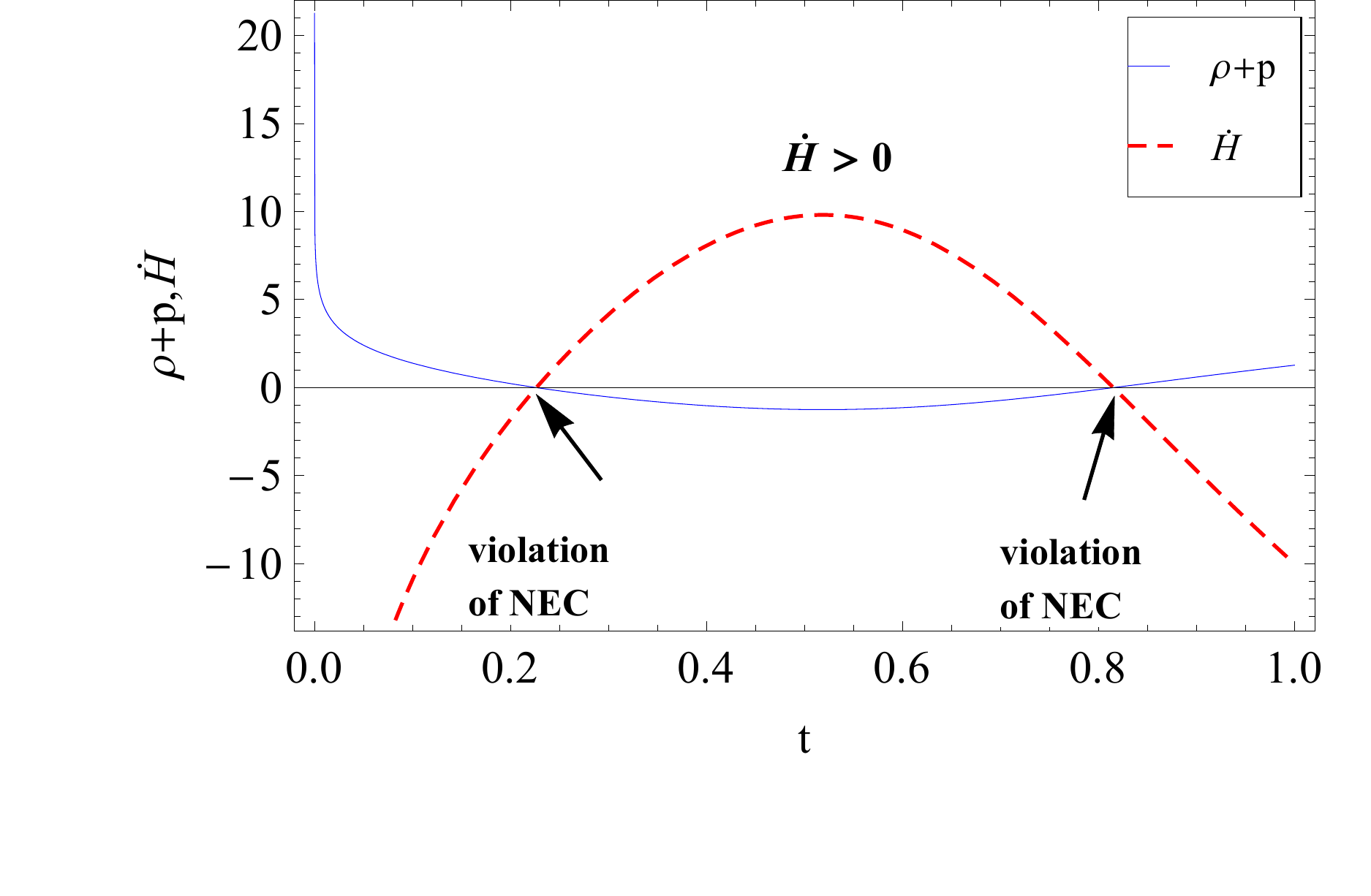} & \includegraphics[width=3in]{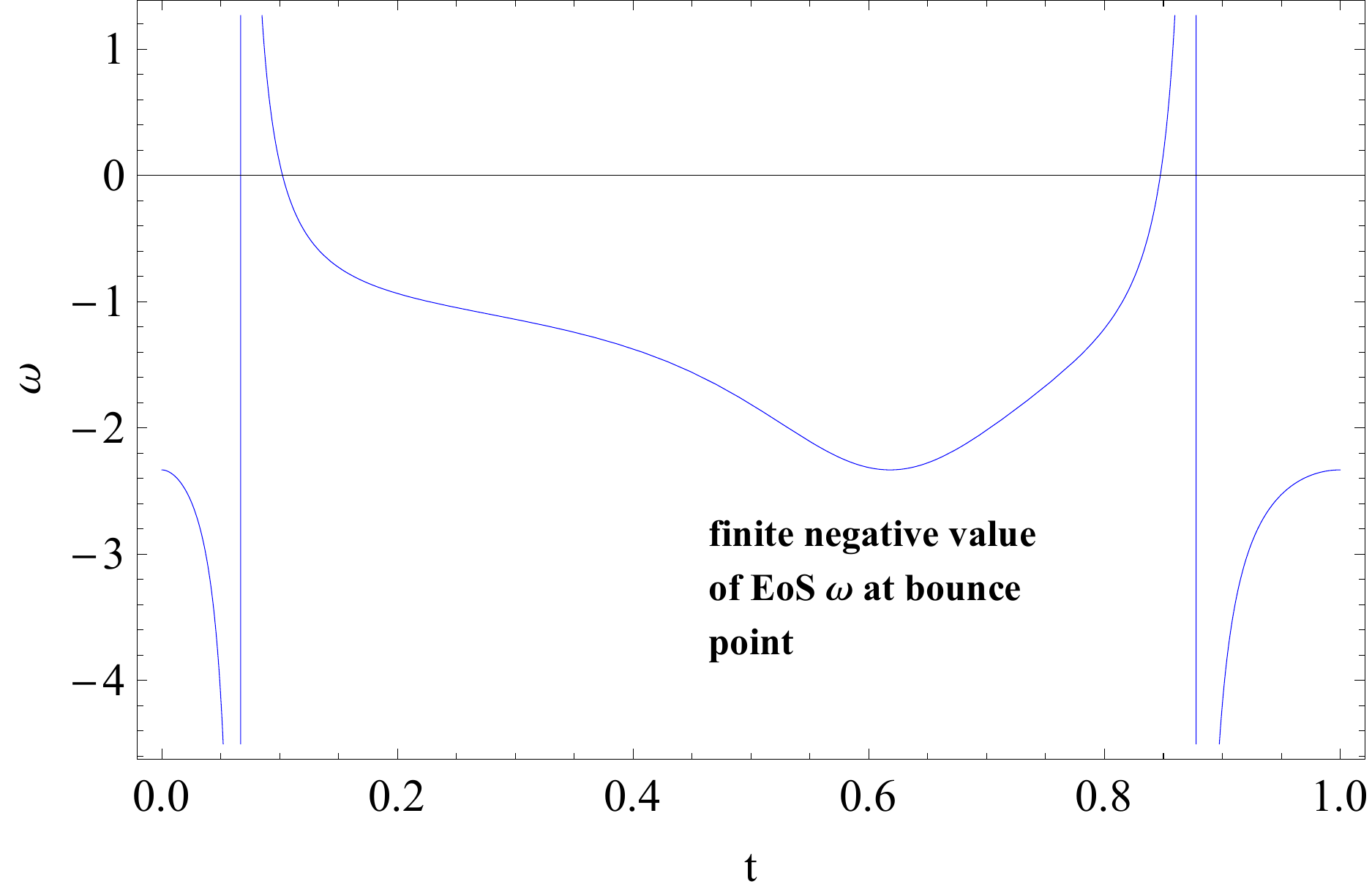}
&  \\
\mbox (a) & \mbox (b) &
\end{array}%
$ }
\end{center}
\caption{\scriptsize (a) The plot of null energy condition (NEC) and $\dot{H}$ \textit{%
vs.} $t$ for $\protect\alpha =-10$ and $\protect\lambda_1 =-1.5\protect\pi $.
(b) The plot of EoS parameter $\protect\omega $ \textit{vs.} $t$ for $%
\protect\alpha =-10$ and $\protect\lambda_1 =-1.5\protect\pi $.}
\end{figure}

From Fig. 4 and 5, we observe that

\begin{itemize}

\item[(i)] the Hubble parameter $ H<0 $ in the interval $ 0<t<0.618  $, $ H>0 $ in the interval $ 0.618<t<1 $ and $ H=0 $ at $ t\simeq0.618 $ for $ \alpha<0 $ (see Fig. 4a),

\item [(ii)] During the contracting universe, the scale factor $a(t)$ is decreasing \textit{ i.e.} $ \dot{a(t)} < 0 $, and the scale factor $ a(t) $ shows increasing pattern, \textit{i.e.} $ \dot{a(t)} > 0 $ during the expanding phase of the Universe. The scale factor of the Universe reaches to a non-zero minimum value $ a\simeq0.348 $ at the transfer point $ t\simeq0.618 $ for $ \alpha<0 $ (see Fig. 4b),

\item [(iii)] $ \dot{H}> 0 $ in the interval $ 0.2295<t<0.8167 $ \textit{i.e.} in the neighborhood of bouncing point at $ t\simeq0.618 $. Therefore the null energy condition (NEC) is violated in same interval (see Fig. 5a),

\item [(iv)] Our obtained model is a Quintom model as the EoS parameter $\omega$ of the matter content undergoes a phase transition from $ \omega<-1 $  to $ \omega>-1 $ in the neighborhood of bouncing point at $ t\simeq0.618 $. Therefore, in this scenario our Universe enters into the hot big bang age after the bounce. \cite{cai7,cai8} (see Fig. 5b).

\end{itemize}

Since all the above criterion are fulfilled by our derived cosmological model. Therefore, we can say that our model is a non-singular bouncing model within FLRW Universe in the background. Moreover,  this model behaves like a Quintom model \cite{fen} which is supported by the observational data \cite{zho}. On the basis of above observations, we can predict that this model is very helpful to study the behaviors of the DE with an EoS parameter  $w >-1$ in the past and  $w <-1$ at present.

\section{\textbf{Scalar field description}}

\qquad In the recent years, the Quintom model have earned a great popularity to study the bouncing cosmological model within GTR. The simplest Quintom model contains two types of scalar fields: one is the quintessencelike and other is the Phantomlike. However, it is not easy to constitute a Quintom model theoretically since $ \omega=-1 $ is not consistent with observations. If we want our model to be consistent with observations, we need to take $ \omega \simeq-1 $. Thus we need $ \dot{\phi}^{2}<<V(\phi) $ \textit{i.e.} the kinetic energy of the scalar field is negligible in comparison to the potential energy.   If $ \omega \simeq-1 $, there are many models to explain acceleration. We can use exactly the same model for inflation. Here, we are interested to study the non-singular bouncing cosmological model using scalar fields in the background of f(R,T) gravity.\\

The Einstein theory of gravity is defined by the following action

\begin{equation}\label{15a}
S=\frac{c^4}{16\pi G}\int R\sqrt{-g}d^{4}x+S_m,
\end{equation}
where $ S_m $ is the action for the quintessencelike and phantomlike scalar 
field denoted by $ S_{m_q} $ and $ S_{m_{ph}} $ respectively. Here we use normalization by taking $ c=1 $.\\

The action for the quintessencelike and phantomlike scalar field are given by 

\begin{equation}\label{15b}
S_{m_q}=\int \left[ -\frac{1}{2}\partial_\mu{\phi}_q\partial^\mu{\phi}_q-V(\phi _q)\right]  \sqrt{-g}d^{4}x,
\end{equation}

and

\begin{equation}\label{15c}
S_{m_{ph}}=\int \left[ \frac{1}{2}\partial_\mu{\phi}_{ph}\partial^\mu{\phi}_{ph}-V({\phi}_{ph} )\right]  \sqrt{-g}d^{4}x,
\end{equation}
respectively. As the scalar field $\phi$ is time dependent, therefore it can be considered as perfect fluid with energy density $ \rho _{\phi } $ and pressure $ p_{\phi} $. We assume that if the scalar field $\phi$ is the only source of DE having potential $V(\phi )$ which interacts with itself, so we can consider energy densities $\rho _{\phi_q }$ , $\rho_{\phi_{ph}}$ and pressures $p_{{\phi}_q}$, $p_{\phi_{ph}}$ for the quintessencelike and phantomlike scalar fields in the framework of FLRW cosmology as

\begin{equation}\label{16}
\rho _{{\phi}_q}=\frac{1}{2} \dot{\phi}_{q}^{2}+ V(\phi _q),\,\,\, p_{{\phi}_q}=\frac{1}{2} \dot{\phi}_{q}^{2}- V(\phi _q),
\end{equation}

\begin{equation}\label{17}
\rho_{\phi_{ph}}=-\frac{1}{2} \dot{\phi}_{ph}^{2}+V(\phi_{ph}),\,\,\,p_{{\phi}_{ph}}=-\frac{1}{2} \dot{\phi}_{ph}^{2}-V(\phi_{ph}).
\end{equation}
Here, the suffixes $ q $ and $ ph $ correspond to quintessencelike and phantomlike scalar 
field respectively. The kinetic energies $ \frac{1}{2} \dot{\phi}_{q}^{2} $, $ \frac{1}{2} \dot{\phi}_{ph}^{2} $ and potential energies $ V(\phi _q) $, $ V(\phi_{ph}) $ of  quintessencelike and phantomlike scalar field correspondence are

\begin{eqnarray}\label{18}
\frac{1}{2}\dot{\phi}^{2}_{q} =\frac{\alpha (1-2t^3+(-1+t-t^3)) ln[-1+\frac{1%
}{t}+t^2]}{2(4\pi+\lambda_1)(1-t+t^3)},
\end{eqnarray}

\begin{eqnarray}\label{19}
\frac{1}{2}\dot{\phi}^{2}_{ph} =-\frac{\alpha[-1+2t^3+(1-t+t^3) ln(-1+\frac{1%
}{t}+t^2)+3\alpha t^2(1-t+t^3) ln(-1+\frac{1}{t}+t^2)^2]}{%
4(2\pi+\lambda_1)(1-t+t^3)},
\end{eqnarray}

\begin{equation}\label{20}
V(\phi _{q})=V(\phi _{ph})=\frac{\alpha \lbrack -1+2t^{3}+(1-t+t^{3})ln(-1+%
\frac{1}{t}+t^{2})(1+3\alpha t^{2}ln(-1+\frac{1}{t}+t^{2}))]}{4(2\pi
+\lambda_1 )(1-t+t^{3})}.
\end{equation}

\begin{figure}[tph]
\begin{center}
{\scriptsize $%
\begin{array}{c@{\hspace{.1in}}cc}
\includegraphics[width=2.3in]{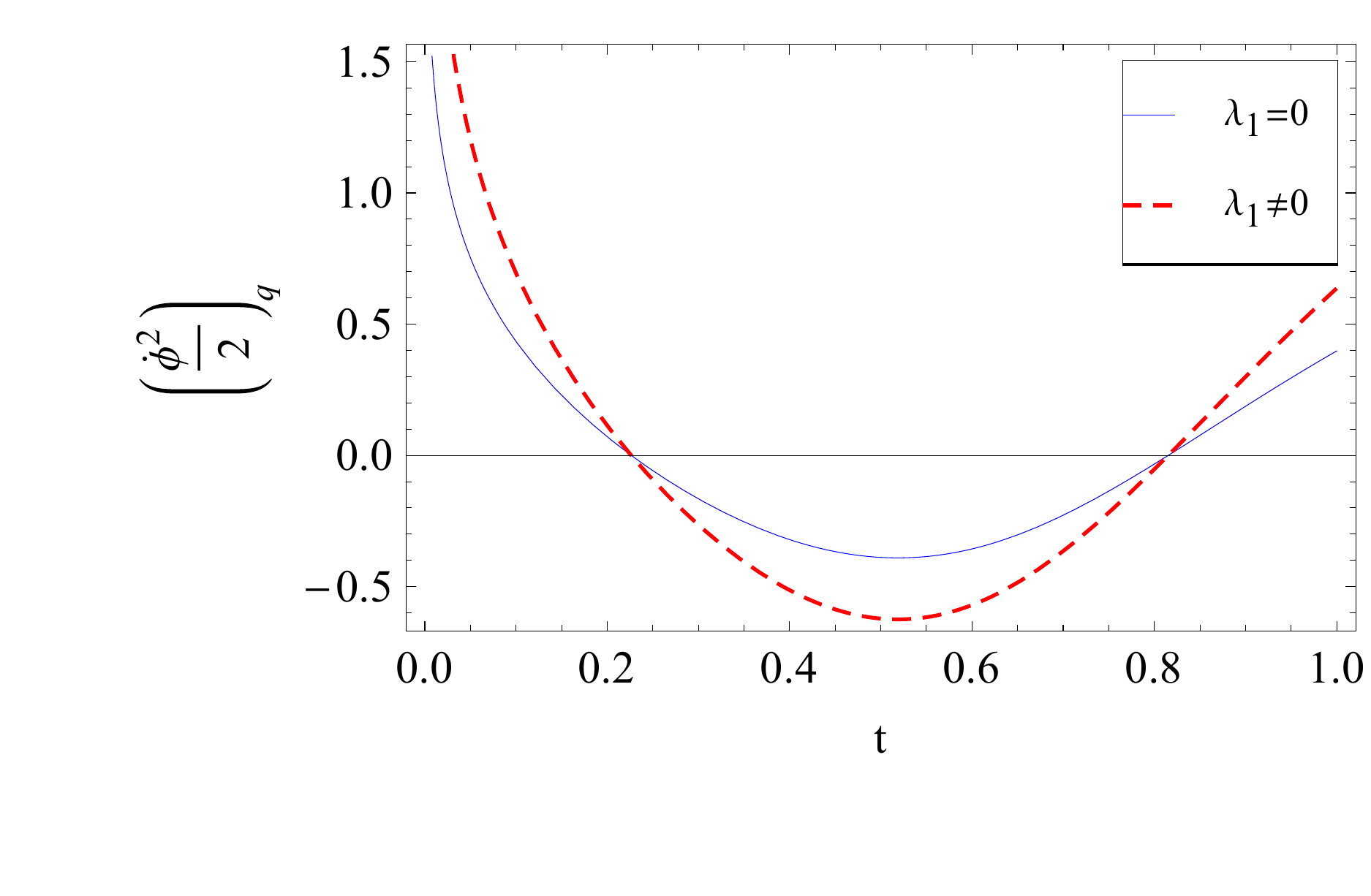} & \includegraphics[width=2.3in]{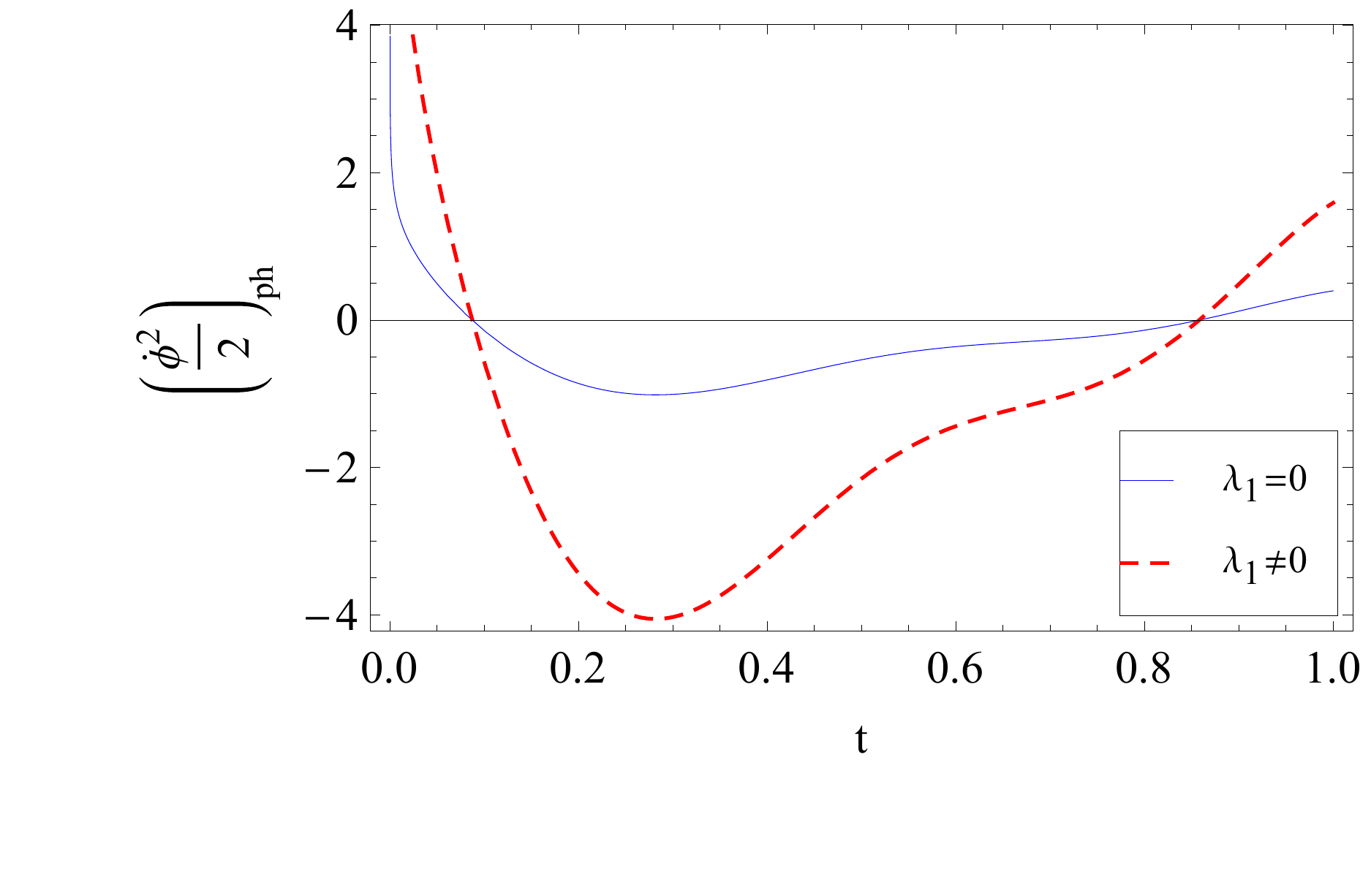}
& \includegraphics[width=2.3in]{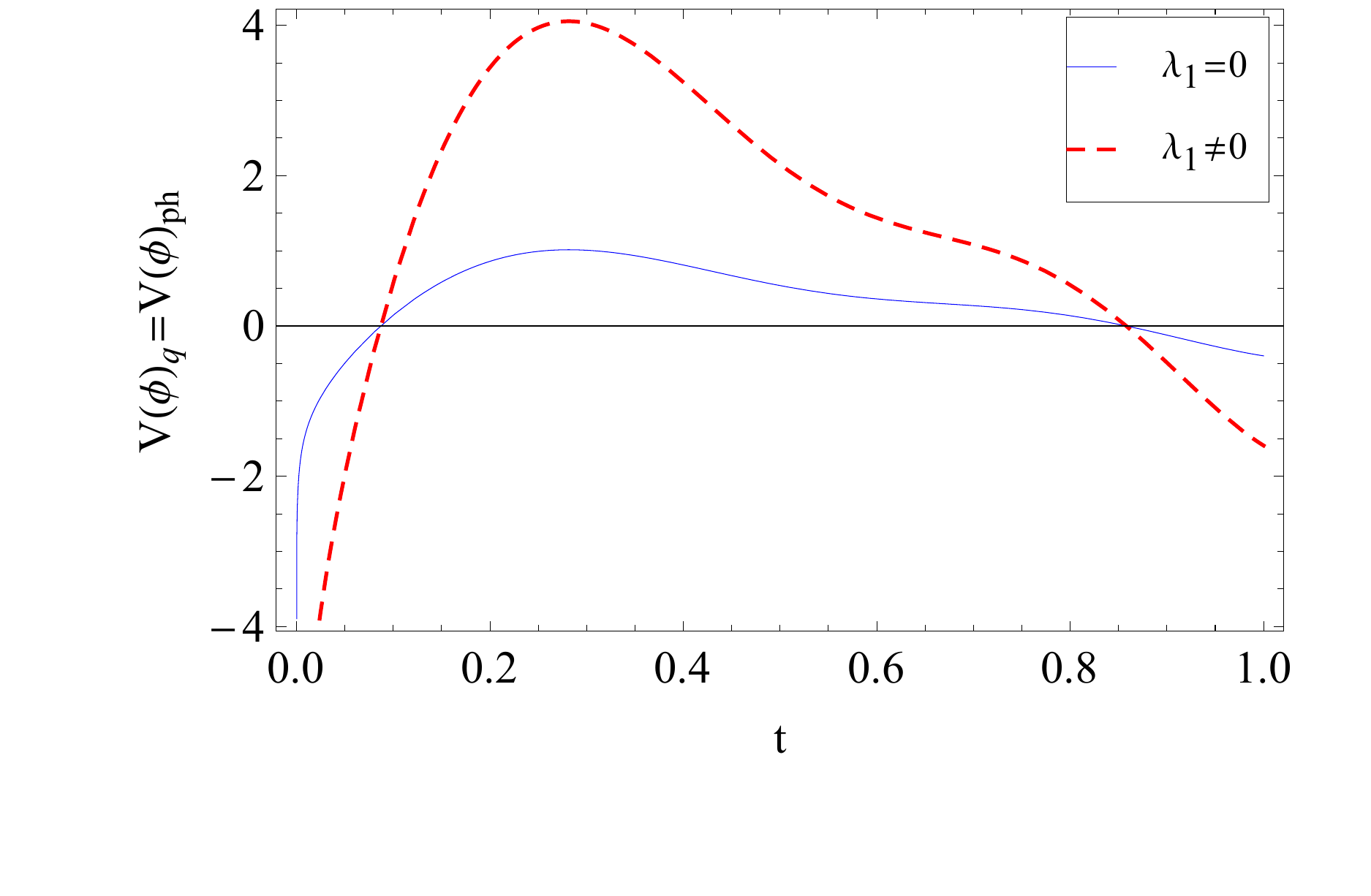} \\
\mbox (a) & \mbox (b) & \mbox (c)\\
\end{array}%
$ }
\end{center}
\caption{\scriptsize (a) The plots of kinetic energy $ \frac{1}{2} \dot{\phi}_{q}^{2} $ \textit{vs.} $t$ for quintessencelike scalar field in $f(R,T)$ gravity and GTR for $\protect\alpha =-10$. (b)  The plots of kinetic energy $ \frac{1}{2} \dot{\phi}_{ph}^{2} $ \textit{vs.} $t$ for phantomlike scalar field correspondence in $f(R,T)$ gravity and GTR for $\protect\alpha =-10$. (c) The plots of potential energy $V(\phi)$ \textit{vs.} $t$ for quintessence and phantomlike scalar field correspondence in $f(R,T)$ gravity and GR for $\protect\alpha =-10$.}
\end{figure}

In Fig. 6a, 6b and 6c, we analyze the variation of kinetic energy  $ \frac{1}{2} \dot{\phi}_{q}^{2} $, $ \frac{1}{2} \dot{\phi}_{ph}^{2} $, potential energy $ V(\phi _q) $ and $ V(\phi_{ph}) $ of quintessencelike  and phantomlike scalar field \textit{w.r.t.} cosmic time $ t $ in $f(R,T)$ gravity as well as in GTR. The negative values of $ \frac{1}{2} \dot{\phi}_{q}^{2} $ and $ \frac{1}{2} \dot{\phi}_{ph}^{2} $ show the DE models due to repulsive force in the interval $ 0.2252<t<0.8151 $ (Fig. 3a) and $ 0.08995<t<0.857 $ (Fig. 3b) in the neighborhood of bouncing point at $ t\simeq0.618 $. From Eq. (\ref{20}), we find that the potential energy $V(\phi _q)$ and $V(\phi_{ph})$ for quintessence and phantomlike scalar field are equal and positive in the interval $ 0.08798<t<0.859 $ in the neighborhood of bouncing point at $ t\simeq0.618 $.

In case of phantomlike and quintessencelike scalar field, the EoS parameters $\omega$ are given by

\begin{equation}\label{20a}
\omega=\frac{p_{\phi_{ph}}}{\rho_{\phi_{ph}}}<-1,\,\,\,\omega=\frac{p_{\phi_q}}{\rho_{\phi_q}}>-1.
\end{equation}

In case of Quintom behavior of the model when the EoS parameter $ \omega $ crosses over the line $ \omega=-1 $
then from Eqs. (\ref{18}), (\ref{19}) and (\ref{20}), we have 

\begin{equation}\label{20b}
\frac{1}{2} \dot{\phi}_{q}^{2}= \frac{1}{2} \dot{\phi}_{ph}^{2},  
\end{equation}
which is the necessary condition to the model having bouncing behavior and is consistent with the results of Cai \textit{et al.} \cite{cai7}. Hence, we conclude that our model is a non singular bouncing model in $f(R,T)$ gravity.

\section{Discussions and Conclusions}

\noindent \qquad In this paper, we have studied the flat FLRW model with a specific form of HP, which is a function of time $t$. Several different forms of HP have already been proposed in literature, but our parametrization possesses some specific features. We have obtained the deterministic solution to EFEs under our parametrization scheme (\ref{9}) by assuming $\lambda_1 =\lambda_2 = \lambda$. Further, we have investigated some restrictions on the model parameters $\alpha $, $\lambda _{2}$, $c$ leading to some cases of expanding Universe ($H>0$) in Table 1. Also, in Table 2, we have found restrictions on model parameters which shows eternal acceleration ($q<0$) . We have examined the physical behavior of the deceleration parameter,  energy density, matter pressure, and the EoS parameter for the model. In order to have a concrete understanding of the bouncing scenario for our parametrization, we have considered a more concise form of the function h(t), given in (\ref{8}) by providing some particular values to the model parameters and have discussed all the necessary conditions for a successful bouncing model. Lastly, we have also discussed the self interacting potential $V(\phi )$ and kinetic energy $\frac{\dot{\phi}^{2}}{2}$ for quintessence and phantom scalar field correspondence in the presence of $f(R,T)$ gravity and compared it with GR by considering scalar field $\phi $ as the source of DE.

\begin{itemize}
\item  In order to study the bouncing nature of the model, various restrictions have been imposed on the model parameters. Under some restrictions, our model shows bouncing behavior at $ t=0 $ and at $ t=\frac{3(1+\lambda_{2})2^{\frac{1}{3}}}{\left( -81c\lambda _{2}^{2}+\sqrt{6561c^{2}\lambda _{2}^{4}-2916\lambda _{2}^{3}(1+\lambda_{2})^{3}}\right) ^{\frac{1}{3}}}+\frac{\left( -81c\lambda _{2}^{2}+\sqrt{6561c^{2}\lambda _{2}^{4}-2916\lambda _{2}^{3}(1+\lambda _{2})^{3}}\right) ^{\frac{1}{3}}}{3\lambda _{2}2^{\frac{1}{3}}} $ (approx.). We have examined the future bounce by varying the model parameters $ c $ and $ \lambda _{2} $ (see Fig. 1a, 1b), and it is observed that the future bounce is delayed by rescaling the model parameters $c$ and $\lambda_2$ (either by increasing $ c $ or decreasing $ \lambda _{2}$). The model parameters have been taken in such a way that some specific features of our proposal could be studied.

\item The model exhibits eternal acceleration throughout the evolution of the Universe with some restrictions on model parameters estimated in Table 2. The physical behavior of EoS parameter under some restrictions on $ H >0$,  mentioned in Table 1 is shown in Fig. 3. In our parametrization of $ H $ in Eq. (\ref{9}), $\omega_{1}$ and $\omega_{4}$ shows transition from perfect fluid ($0<\omega<1$) to DE region ($\omega<0$) at $t\simeq 0.825$ and $t\simeq0.26$ respectively. At this time the model represents dust Universe ($p=0$), whereas in all the other cases, the model represents DE only. The EoS parameter $\omega$ approaches to quintom line in all the cases at late time.

\item In order to have a concrete understanding of our proposed parametrization of $ H $ in Eq. (\ref{9}), and to explain the bouncing process more precisely, we generate a new parametrization of $ H $ in Eq. (\ref{15}) in a specific form with some particular values of the model parameters $ c $ and $ \lambda_2 $. The bouncing scenario can be accomplished for both negative and positive scaling constant $\alpha$ but here to get an expanding Universe from the prior period of contraction, we have chosen $\alpha$ with negative value1. Some plots have been presented in order to achieve necessary conditions for a successful bouncing model (see Fig. 4, 5). In the parametric form of $ H $ (\ref{15}), the bouncing point is attained at cosmic time $t\simeq0.618 $, which leads to a minimum, non vanishing value of scale factor $ a(t) $ (see Fig. 4a, 4b).

\item For the specific form of $ H $ in Eq. (\ref{15}), the model depicts the bouncing process (\textit{i.e.} expansion before bounce and contraction after bounce for $\alpha >0$, and contraction before bounce and expansion after bounce for $\alpha <0$). The first derivative of Hubble parameter $ \dot{H}>0 $ during the period $t \in (0.2295 ,0.8167)$ leads the violation of null energy condition (NEC), which is the compelling condition for a bouncing scenario in our model (see Fig. 5a). Our model is a Quintom model in which the EoS parameter of the matter content $ \omega $ transits from phantom phase $ \omega<-1 $  to quintessence phase $ \omega>-1 $ in the neighborhood of bouncing point at $ t\simeq0.618 $. Therefore, we can see that the Universe enters into the hot big bang era after the bouncing (see Fig. 5b).

\item In section 5, we have discussed the quintessence-like and phantom-like scalar fields in $ f(R,T) $ gravity and GTR for the parameterization of $ H $ stated in Eq. (\ref{15}). We have observed that both K.E. and P.E. exhibit similar pattern with different scaling in $ f(R,T) $ gravity and GTR (see Fig 6). In this case, our model behaves as Quintom model provided the condition given in Eq. (\ref{20b}) is satisfied, which is the necessary condition to the model having bouncing behavior. This condition is also consistent with the results of Cai \textit{et al.} \cite{cai7}. Therefore, we say that our model is a non singular bouncing model in $f(R,T)$ gravity. 

\item Thus we conclude that the bouncing scenario of the cosmological model have been discussed in $ f(R)$ gravity where the scale factor is taken in the forms of exponential and power-law, Gauss-Bonnet gravity, $ f(G)$ gravity where the Gauss-Bonnet invariant is taken as $ G $, and $ f(T) $ gravity where $ T $ is the torsion scalar in the teleparallelism but we have studied a non-singular bouncing cosmological model in $ f(R,T) $ gravity within a flat FLRW background metric with a specific parametrization of the Hubble parameter which is the main difference of my research work and may also be useful for further investigation.

\end{itemize}

\vskip0.2in \textbf{\noindent Acknowledgements } The authors express their thanks to Centre for Theoretical Physics, Jamia Millia Islamia, New Delhi, India for some fruitful discussions with Prof. M. Sami and Prof. S. G. Ghosh for providing necessary facilities to complete the work.  The author JKS expresses his thanks to Department of Mathematical Sciences, University of Zululand, Kwa-Dlangezwa 3886, South Africa and Department of Mathematics, Statistics and Computer Sciences, University of KwaZulu-Natal, Westville 4001, South Africa for some fruitful discussions with Prof. A. Beesham and Prof. S. D. Maharaj, their financial supports and providing necessary facilities as well as hospitalities where a part of the work has been completed. Author SKJP wishes to thank NBHM (DAE) for financial support through post-doctoral research fellowship. Moreover, the work of KB was supported by the JSPS KAKENHI Grant Number JP25800136 and Competitive Research Funds for Fukushima University Faculty (17RI017). Authors also express their thanks to the referee for his valuable comments and suggestions.

\end{document}